\documentclass[12pt]{article}

\usepackage{scicite}

\usepackage{times}
\usepackage[labelfont=bf]{caption}

\usepackage{amsmath}
\usepackage{amsfonts}
\usepackage{amssymb}
\usepackage{graphicx}
\usepackage{pdfpages}

\newenvironment{sciabstract}{%
\begin{quote} \bf}
{\end{quote}}

\title{Molecular beam homoepitaxy of N-polar AlN: enabling role of Al-assisted surface cleaning}

\author
{Zexuan Zhang,$^{1\ast}$ Yusuke Hayashi,$^{2}$ Tetsuya Tohei,$^{2}$ Akira Sakai,$^{2}$ \\Vladimir Protasenko,$^{1}$ Jashan Singhal,$^{1}$ Hideto Miyake,$^{3,4}$ \\Huili Grace Xing,$^{1,5,6}$ Debdeep Jena,$^{1,5,6}$ and YongJin Cho$^{1\ast}$\\
\\
\normalsize{$^{1}$School of Electrical and Computer Engineering, Cornell University, }\\
\normalsize{Ithaca, New York 14853, USA}\\
\normalsize{$^{2}$Graduate School of Engineering Science, Osaka University, }\\
\normalsize{1-3 Machikaneyama-cho, Toyonaka, Osaka 560-8531, Japan}\\
\normalsize{$^{3}$Graduate School of Engineering, Mie University, }\\
\normalsize{1577 Kurimamachiya-cho, Tsu, Mie 514-8507, Japan}\\
\normalsize{$^{4}$Graduate School of Regional Innovation Studies, Mie University, }\\
\normalsize{1577 Kurimamachiya-cho, Tsu, Mie 514-8507, Japan}\\
\normalsize{$^{5}$Department of Materials Science and Engineering, Cornell University, }\\
\normalsize{Ithaca, New York 14853, USA}\\
\normalsize{$^{6}$Kavli Institute at Cornell for Nanoscale Science, }\\
\normalsize{Ithaca, New York 14853, USA}\\
\\
\normalsize{$^\ast$To whom correspondence should be addressed; E-mail:  zz523@cornell.edu}\\
\normalsize{and yongjin.cho@cornell.edu}\\
}

\date{}

\begin{document} 


\baselineskip24pt


\maketitle



\begin{sciabstract}
  N-polar aluminum nitride (AlN) is an important building block for next-generation high-power RF electronics. We report successful homoepitaxial growth of N-polar AlN by molecular beam epitaxy (MBE) on large-area cost-effective N-polar AlN templates. Direct growth without any \textit{in-situ} surface cleaning leads to films with inverted Al-polarity. It is found that Al-assisted cleaning before growth enables the epitaxial film to maintain N-polarity. The grown N-polar AlN epilayer with its smooth, pit-free surface duplicates the structural quality of the substrate as evidenced by a clean and smooth growth interface with no noticeable extended defects generation. Near band-edge photoluminescence peaks are observed at room temperature on samples with MBE-grown layers but not on the bare AlN substrates, implying the suppression of non-radiative recombination centers in the epitaxial N-polar AlN. These results are pivotal steps towards future high-power RF electronics and deep ultraviolet photonics based on the N-polar AlN platform.  
\end{sciabstract}


\section*{Introduction}

High electron mobility transistors (HEMTs) built on wide-bandgap semiconductor material platforms such as III-nitrides are leading contenders in high-power mm-wave electronics\cite{hickman2021next,romanczyk2020w,hamza2020review}. Compared with their metal-polar counterparts, nitrogen-polar GaN-based HEMTs allow for the simpler formation of low-resistance contacts due to the absence of a top barrier and stronger carrier confinement thanks to the inherent back barriers\cite{wong2007low,wong2008n}. Current state-of-the-art performance has been achieved using N-polar GaN/AlGaN HEMTs with output powers above 8 W/mm at up to 94 GHz\cite{romanczyk2020w}. The performance of N-polar III-N HEMTs can potentially be further improved with binary AlN buffer layers\cite{lemettinen2018n,ito2020growth}. Because of its large bandgap (6 eV) and high thermal conductivity ($\sim$ 340 W/mK), AlN provides an unmatched combination of high electrical resistivity and thermal conductivity in the nitride semiconductor family\cite{hickman2019high,hickman2021next}. As a result, incorporating free-standing N-polar AlN as the buffer layer for N-polar III-N HEMTs has the advantages of enhanced thermal management and a maximized conduction band offset to help reduce buffer leakage and short channel effects\cite{hickman2021next,lemettinen2018n,ito2020growth,wong2008n,lemettinen2019nitrogen}. In addition, substituting the AlGaN buffer layer with AlN can induce a higher density of two-dimensional electron gas (2DEG), suppress alloy scattering, and has the potential to further boost the conductivity of the 2DEG channel\cite{smorchkova2001aln,lemettinen2018n,ito2020growth,wong2008n}. Moreover, III-N HEMTs on AlN can take advantage of the unprecedented level of integration in nitride electronics provided by the AlN platform\cite{hickman2021next}. 

The first step to achieve N-polar III-N HEMTs based on the AlN platform is the epitaxial growth of high-quality N-polar AlN. N-polar AlN has been synthesized on different \emph{foreign} substrates such as Si, SiC and sapphire using techniques such as metal-organic vapor phase epitaxy (MOVPE), sputtering and molecular beam epitaxy (MBE)\cite{dasgupta2009growth,ledyaev2014n,lemettinen2018movpe,isono2020growth,okumura2012growth,shojiki2021reduction,liu2021polarity,keller2006effect}. Among these, optimal conditions for the MOVPE growth of N-polar AlN have been developed on C-face SiC substrates with an intentional miscut of 1\textdegree~  to achieve a smooth surface free of hexagonal hillocks and step bunching\cite{lemettinen2018movpe}. This further led to the recent demonstration of N-polar AlGaN/AlN polarization-doped field-effect transistors (PolFETs)\cite{lemettinen2019nitrogen}. Yet, to maximize the performance of such N-polar AlN-based devices, the development of homoepitaxial growth technique on N-polar AlN substrate is highly desired. Reports on homoepitaxy of N-polar AlN, however, are rare. Although N-polar AlN homoepitaxy on single-crystal AlN substrates has been recently demonstrated by MOVPE\cite{isono2020growth}, successful N-polar AlN homoepitaxy by MBE has not been reported yet. 

In this work, we report the MBE homoepitaxy of N-polar AlN films on N-polar AlN templates. By comparing two samples with and without \emph{in-situ} Al-assisted surface cleaning of the substrate, we find Al-assisted surface cleaning to be crucial for achieving N-polarity of the MBE grown AlN epilayers. The MBE grown N-polar AlN is electrically insulating with a smooth pit-free surface and a high structural quality. No disordered interfacial layer or generation of extended defects are detected at the growth interface. In addition, near band-edge photoluminescence emission, absent from the bare substrate, is observed at room temperature on samples with MBE grown layers, suggesting superior optical quality of the MBE-grown N-polar AlN.

\section*{Results}


Two samples with AlN layers grown by MBE on N-polar AlN templates are compared in this work. Except for the presence/absence of \emph{in-situ} Al-assisted surface cleaning before growth (as discussed later), the layer structures and growth conditions for the two samples are nominally identical. The inset of Fig.~1B shows a schematic of the sample structures used in this study. The difference between the two samples lies in the \emph{in-situ} surface cleaning before the MBE growth. For sample A, no intentional \emph{in-situ} cleaning was performed, whereas for sample B, Al-assisted surface cleaning was employed before the MBE growth. The Al-assisted surface cleaning consists of multiple cycles of Al adsorption and desorption, similar to earlier reports on Al-polar AlN substrates\cite{cho2020molecular,lee2020surface}. The substrate was first heated up to a thermocouple temperature of 1060 \textdegree C without nitrogen gas flow. During each Al adsorption/desorption cycle, the substrate was exposed to an Al flux with a beam equivalent pressure (BEP) of $\sim$ 6$\times$10$^{-7}$~Torr for 30~s. The Al shutter was then closed long enough for all of the deposited Al to desorb. 

This Al adsorption and desorption process was clearly observed via the time evolution of the reflection high-energy electron diffraction (RHEED) intensity. Figure~1A shows the variation of the RHEED intensity during the first five Al-assisted cleaning cycles. As can be seen from Fig.~1A, the RHEED intensity drops when Al is deposited and gradually increases, eventually saturating, when it desorbs. Similar behavior was also observed for Al-polar AlN substrates\cite{cho2020molecular,lee2020surface}. The time for the deposited Al to completely desorb from the surface (monitored by the saturation of RHEED intensity) monotonically \emph{decreases} with increasing number of Al-assisted cleaning cycles. The reason for the shortening of the Al desorption time is the gradual removal of surface oxide by Al metals, as has been explained in our previous work\cite{cho2020molecular}. During each Al-assisted cleaning cycle, deposited Al metal reacts with the surface oxide to produce a volatile suboxide, which evaporates at high substrate temperature\cite{cho2020molecular,lee2020surface}. Starting at a thermocouple temperature of 1060 \textdegree C, Al-assisted cleaning cycles were repeated until the desorption time dropped below $\sim$ 50~s, at which point the substrate temperature was lowered by 30 \textdegree C; this process was then repeated until the substrate thermocouple temperature reached 940 \textdegree C. This repeated temperature lowering was carried out to increase the lifetime of Al adatoms on the surface, providing them enough time to react with any residual surface oxides. In order to guarantee thorough surface oxide removal, a total of one hundred Al-assisted cleaning cycles were performed. Figure~1B shows the RHEED intensity vs time during the last five Al-assisted cleaning cycles at a lowered substrate temperature of 940 \textdegree C. Almost no change in the evolution of the RHEED intensity are observed during these cycles and we use this as an indicator of the surface being sufficiently cleaned\cite{cho2020molecular}. It is interesting to note that the two-step adsorption/desorption process, which is characterized by a sharp change in the slope of RHEED intensity vs time (e.g., see Fig.~1 in Ref. 19) and observed in an Al-polar AlN substrate due to a clear transition between adlayer and droplet formation/desorption\cite{cho2020molecular,lee2020surface}, is not observed on the N-polar AlN substrate, indicating a relatively smaller diffusion length of Al adatoms on the N-polar AlN substrate surface.

The evolution of the RHEED patterns of both samples (viewed along the AlN $<$11$\bar{2}$0$>$ azimuth) during the growth are displayed in Fig.~2. For sample A, the RHEED pattern was slightly diffuse with faint streaks prior to the growth (Fig.~2A). At the nucleation stage, the RHEED pattern became completely diffuse (Fig.~2B), indicating a very high level of surface crystalline disorder. As growth proceeded, the RHEED pattern started to brighten and streaks gradually recovered. Figure~2C shows the RHEED pattern by the end of the growth after desorption of excess Al droplets at 970 \textdegree C. The bright and streaky RHEED pattern suggests a smooth surface. In contrast, the RHEED pattern for sample B before epitaxial growth (after Al-assisted cleaning) was bright and streaky (Fig.~2D). Such RHEED pattern persisted throughout the entire growth, as shown in Figs.~2 (E and F). No considerable change in the RHEED pattern was observed during cooling down the substrates to room temperature for both samples.  

Examining the surface morphology of the two samples by atomic force microscopy (AFM), although sample A has a smooth surface morphology with a low root-mean-square (rms) roughness of ~0.6 nm in a 10$\times$10 $\mu$m$^2$, it has pits and trenches on the surface (Figs.~3, A and B). Apart from these features, clear atomic steps are observed, suggesting a step-flow growth mode enabled by Al-rich growth conditions. On the other hand, sample B is very smooth with an rms roughness as low as 0.3~nm in a 10$\times$10~$\mu$m$^2$ region (Fig.~3C). In addition, the 2$\times$2 $\mu$m$^2$ AFM scan in Fig.~3D shows the presence of smooth and parallel atomic steps. No visible hexagonal hillocks or surface pits were observed. The origin of the surface pits in sample A could be attributed to relatively high-density contaminants including oxides which are presumably present on the substrate surface due to the lack of any \emph{in-situ} surface cleaning. Similar pits have been found in films grown on N-polar GaN substrate with high density C impurities on the substrate surface\cite{wurm2020growth}. 

To determine the polarity of AlN layers, KOH etching is widely used, due to the significantly different etch rates for Al-polar and N-polar nitride surfaces\cite{lemettinen2018movpe,hong2014investigation,kirste2013polarity}. Specifically, Al-polar AlN exhibits defect-selective etch behavior by KOH with hexagonal pits generated around dislocations\cite{lemettinen2018movpe,lu2008microstructure}. In contrast, N-polar AlN can be etched by KOH with a much higher etch rate, with hexagonal pyramids bounded by more chemically stable \{1$\bar{1}$0$\bar{1}$\} crystallographic planes emerging after etching\cite{guo2015koh,shojiki2021reduction,lemettinen2018movpe}. Figure~4 shows the surface morphologies of both samples after etching in 50 wt\% KOH aqueous solution at room temperature for 10 minutes. Pits with a density of $\sim4\times10^7$/cm$^2$ were observed in a 5$\times$5 $\mu$m$^2$ AFM scan on sample A shown in Fig.~4A. A zoomed-in 0.5$\times$0.5~$\mu$m$^2$ scan near a pit (the boxed region in Fig.~4A) further reveals its hexagonal shape (Fig.~4B) with a depth of $\sim$ 120 nm (measured by a linescan along the white line). A schematic of sample A after KOH etch is shown in Fig.~4C. As mentioned earlier, such morphology after KOH etch is a signature of Al-polar AlN\cite{lemettinen2018movpe,lu2008microstructure}. 
In sharp contrast, sample B exhibits hexagonal pyramids after KOH etch, indicative of N-polarity, with a density of $\sim2\times10^7$/cm$^2$, as can be seen in Figs.~4 (D and E). A linescan along the white line in Fig.~4E measures the height of the hexagonal pyramid to be $\sim$ 150 nm. Figure~4F shows a schematic of sample B after etching in KOH. 
To further confirm the polarity of both samples, X-ray diffraction (XRD) and Raman spectroscopy were performed. Even after KOH etch, strong AlN peaks (marked by black dashed lines) with intensities comparable to those measured before KOH etch were seen on sample A in both XRD and Raman spectra (Figs.~5, A and B), indicating that the AlN film was not substantially etched by KOH and confirming that the film is Al-polar. For sample B, on the other hand, the AlN peaks in both the XRD and Raman spectra almost completely vanish after KOH etch, as indicated by the black arrows in Figs.~5 (C and D), verifying that the epitaxial film maintained the polarity of the N-polar substrate.

Cross-sectional scanning transmission electron microscopy (STEM) measurements were performed to study the atomic structure and directly probe the polarities of the MBE-grown AlN films in both samples. Figure~6A shows a bright-field (BF) STEM overview image of the cross section of sample A. An obvious interface structure marked by the white notches in Fig.~6A is seen between the sputtered N-polar AlN substrate and the MBE-grown layer, indicating that the AlN is structurally discontinuous across the interface region. In addition, the considerable image contrasts in the MBE-grown layer marked by the black triangles in Fig.~6A, which are absent in the substrate, are considered to be due to strain field from the extended defects generated near the growth interface during the MBE growth. These defects are further identified to be \emph{a}-type dislocations, based on the bright-field transmission electron microscopy (TEM) images shown in the supplemental information (Fig.~S2). Figures~6 (B to D) show the magnified high angle annular dark field-STEM (HAADF-STEM) images of the corresponding regions marked by the black squares in Fig.~6A: Fig.~6B is taken close to the MBE-grown AlN surface, Fig.~6C shows the interface between sputtered and MBE-grown AlN, and Fig.~6D corresponds to the sputtered AlN/sapphire interface. Across a well-defined inversion domain boundary between the white dash lines in Fig.~6C close to the growth interface, the AlN polarity is seen to be inverted: from the N-polarity in the substrate (Fig.~6D) to Al-polarity in the MBE-grown layer (Fig.~6B). Interestingly, this polarity inversion boundary shares a similar microstructure as the one previously reported in sputtered AlN films\cite{akiyma2019structural}. On the contrary, the interface between the MBE-grown layer and the substrate in sample B is not visible in the BF-STEM overview image shown in Fig.~6E, suggesting a high level of structural continuity between the epilayer and the substrate across the interface. Unlike sample A, no sharp image contrast was detected in the MBE-grown layer across the STEM observation area. Moreover, with Al-assisted cleaning before growth, the AlN layer in sample B maintains the polarity of the N-polar substrate, as evidenced by the magnified HAADF-STEM images taken within the MBE-grown AlN layer (Fig.~6F) and the substrate (Fig.~6H). As a result, no polarity inversion boundary is detected between the AlN layer and the substrate, as shown in Fig.~6G. By comparing the atomic structures of samples A and B, it is concluded that Al-assisted cleaning before growth is crucial to achieve a smooth interface and prevent polarity inversion during MBE homoepitaxy. 

It is very likely that surface impurities such as oxygen contribute to the polarity inversion of sample A. In fact, oxygen has been found to play an important role in the polarity inversion from N-polar to Al-polar during AlN growth by other growth techniques including MOVPE and sputtering\cite{stolyarchuk2018intentional,liu2021polarity,shojiki2021reduction}. For example, the polarity of AlN grown on oxygen-plasma treated N-polar AlN surface was found to be Al-polar\cite{stolyarchuk2018intentional}. Besides, the atomic structure of the inversion domain boundary in Fig.~6C resembles the
oxide inversion boundary with Al vacancies formed during sputtering deposition of AlN\cite{akiyma2019structural}. Therefore, \emph{in-situ} Al-assisted cleaning is believed to be effective in deoxidizing N-polar AlN substrates and hence preventing polarity inversion during MBE growth.

Now we move to the structural and optical characterization of the MBE-grown N-polar AlN epilayer (sample B). The structural quality was evaluated by X-ray rocking curves (XRCs), i.e., $\omega$-scans. Figures~7 (A and B) show the measured XRCs of sample B across the symmetrical (0002) and skew-symmetrical (10$\bar{1}$2) reflections, respectively. The full width at half of maximums (FWHMs) of the (0002) and (10$\bar{1}$2) peaks were extracted to be 14 and 380~arcsec, respectively. These values are very close to those [10/350 arcsec for (0002)/(10$\bar{1}$2) peak] measured on the AlN template substrates used in this study\cite{shojiki2021reduction}, suggesting high-quality homoepitaxial growth of the N-polar AlN without noticeable additional generation of  structural defects (see Fig.~6E). 

Figure~7C compares the room-temperature photoluminescence (PL) spectra of sample B and a bare substrate (after subtraction of background from sapphire) near the band-edge of AlN. While no near band-edge emission peak was detected on the bare AlN template, two emission peaks close to the band-edge of AlN were clearly observed on sample B. The emission peak with higher intensity is located at a photon energy of 5.98~eV, which is very close to the reported room temperature free exciton emission line ($\sim$ 5.96~eV) of Al-polar AlN epilayers and bulk AlN crystals\cite{li2002band,feneberg2010high}, whereas the other emission peak at 5.84~eV likely originates from the electron-hole plasma recombination ($\sim$ 5.83~eV on bulk AlN crystals \cite{feneberg2010high}). However, future temperature and excitation power dependent PL measurements are needed to uncover the origin of the PL peaks. Nevertheless, the observation of clear near band-edge PL emission from sample B, which is absent from the bare AlN template, indicates the superior optical quality of the MBE-grown N-polar AlN layer. 



\section*{Discussion}

MBE homoepitaxy of N-polar AlN is achieved on N-polar AlN templates. The \emph{in-situ} Al-assisted surface cleaning before MBE growth is found to be critical in preventing polarity inversion. The MBE-grown N-polar AlN, having a very smooth surface with parallel atomic steps, maintains the high structural quality of the substrate with no noticeable structural distortion or generation of dislocations at growth interface. The superior optical quality of N-polar AlN epilayer is further revealed by the observation of clear room temperature near band-edge PL emission. These results suggest the significant potential of MBE homoepitaxy for preparation of electronic-grade and optical-grade N-polar AlN, and are important milestones towards N-polar RF electronics and deep-UV photonics based on the AlN platform. \newline


\section*{Materials and Methods}

The samples in this study are prepared using molecular beam epitaxy (MBE) in a Veeco GENxplor MBE system equipped with a standard effusion cell for Al and a radio frequency plasma source for active N species. KSA Instruments reflection high-energy electron diffraction (RHEED) apparatus with a Staib electron gun operating at 14.5~kV and 1.45~A was used to \emph{in-situ} monitor the growth front. The substrates used in this study are $\sim$ 160 nm thick N-polar AlN$/$\textit{c}-plane sapphire templates grown by sputtering followed by high-temperature face-to-face annealing. Details about the preparation of N-polar AlN templates can be found elsewhere\cite{shojiki2021reduction}. This cost-effective growth method can produce large-area N-polar AlN templates. After \emph{ex-situ} cleaning in acetone, isopropyl alcohol and deionized water (each for 10 minutes), AlN templates with an area of 1~cm $\times$ 1~cm were mounted in indium-free holders, loaded into the MBE system and outgassed at 200 \textdegree C for 8 h. $\sim$ 300~nm AlN layers were then grown under Al-rich condition at a substrate thermocouple temperature of 940 \textdegree C with an Al beam equivalent pressure (BEP) of $\sim$ 7$\times$10$^{-7}$~Torr and nitrogen plasma operating at 200~W with N$_2$ gas flow rate of 1.95~sccm. After growth, excess Al droplets were desorbed \emph{in-situ} at an elevated thermocouple temperature of 970 \textdegree C before unloading. 

The surface morphologies of the grown samples were characterized by atomic force microscopy (AFM) in an Asylum Research Cypher ES setup. X-ray diffraction (XRD) using a Panalytical XPert Pro setup at 45~kV and 40~mA with the Cu K$\alpha$1 radiation (1.5406~Å) as well as Raman spectroscopy using a 532~nm laser confocal microscope equipped with an 1800~mm$^{-1}$ diffraction grating were also employed for structural characterization. The microstructure of the samples were studied by cross-sectional transmission electron microscopy (TEM) using a JEOL JEM-2100 instrument working at 200~kV. Cross-sectional scanning transmission electron microscopy (STEM) measurements were further performed to directly probe the polarity of the AlN layers using a JEOL ARM-200F system at an accelerating voltage of 200 kV. Prior to the (S)TEM characterization, thin cross-sectional specimens were prepared using a FEI Versa™ 3D DualBeam™ focused ion beam (FIB). Finally, deep ultraviolet photoluminescence (PL) spectroscopy was used to probe the optical quality as well as optical transitions in the N-polar AlN epilayer. Samples were excited from the top using a pulsed ArF excimer laser excitation at 193~nm with 2~mJ energy and a repetition rate of 100~Hz. The emitted light was collected from the side of the samples. 

\section*{Supplementary Materials}
Supplementary Text\\
Figs. S1 to S2\\
References \textit{(30)}

\bibliography{sciadvbib}
\nocite{romano1996inversion}
\bibliographystyle{ScienceAdvances}

\noindent \textbf{Acknowledgements:} 
%
We are indebted to Ryan Page for critical reading of the manuscript.\\
\noindent \textbf{Funding:} The authors at Cornell University acknowledge financial support from the Cornell Center for Materials Research (CCMR)—a NSF MRSEC program (No. DMR-1719875); ULTRA, an Energy Frontier Research Center funded by the U.S. Department of Energy (DOE), Office of Science, Basic Energy Sciences (BES), under Award No. DE-SC0021230; and AFOSR Grant No. FA9550-20-1-0148. This work uses the CESI Shared Facilities partly sponsored by NSF No. MRI DMR-1631282 and Kavli Institute at Cornell (KIC). Y. H. acknowledges funding support from JSPS KAKENHI (grant number 19K15045). H. M. acknowledges funding support from MEXT “Program for Building Regional Innovation Ecosystems.” \\
\noindent \textbf{Author Contributions:} Z.Z., H.G.X., D.J. and Y.C. conceived the research. Z.Z. grew the samples by MBE, conducted the AFM, XRD, Raman spectrosocpy and KOH etching experiments with help from Y.C..  Y.H., T.T. and A.S. performed (S)TEM studies. Y.H. and H.M. performed the growth of the large-area N-polar AlN templates. Z.Z. and V.P. conducted PL measurements. Z.Z., D.J. and Y.C. wrote the manuscript with contribution from all authors.\\
\noindent \textbf{Competing Interests:} The authors declare that they have no competing interests.\\
\noindent \textbf{Data and materials availability:} All data needed to evaluate the conclusions in the paper are present in the paper and/or the Supplementary Materials.


\clearpage

\begin{figure}
\includegraphics[width=0.7 \textwidth]{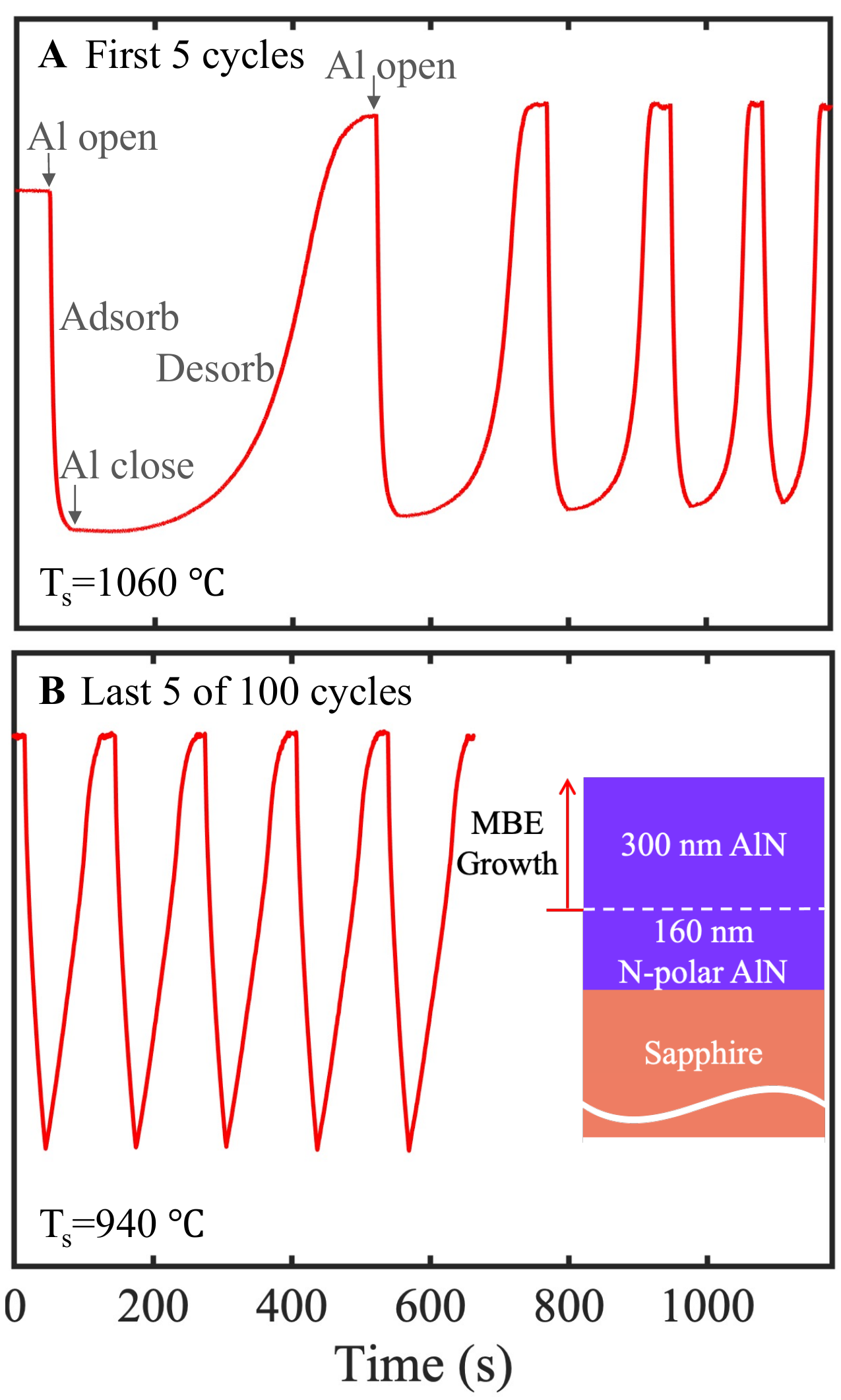}
\centering
\caption{\textbf{RHEED intensity evolution during Al-assisted cleaning}. (\textbf{A}) During the first five cycles at 1060 \textdegree C and (\textbf{B}) during the last five cycles at 940 \textdegree C. Inset: schematic of the sample structures in this study.}
\end{figure}

\clearpage

\begin{figure}
\includegraphics[width=1 \textwidth]{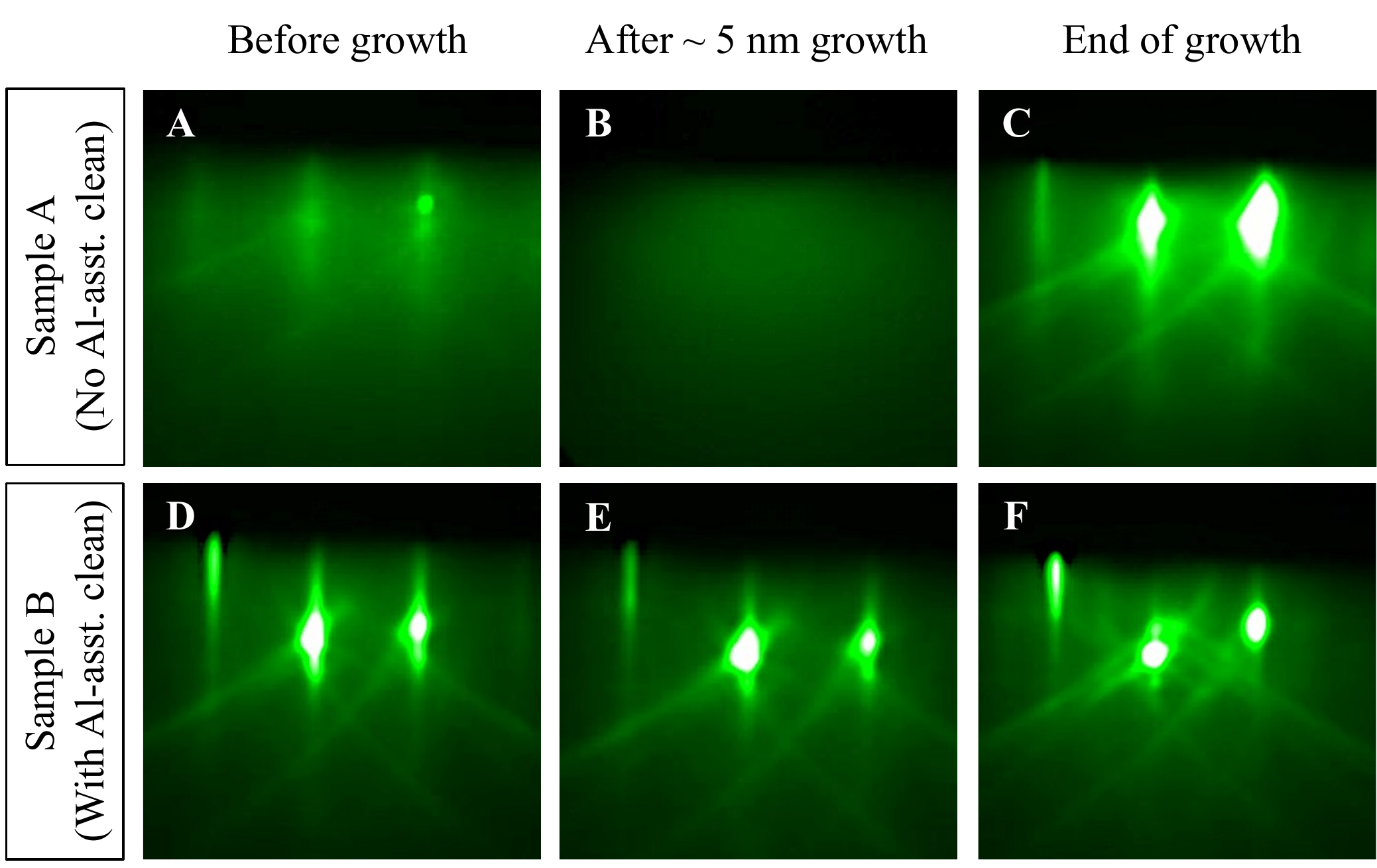}
\centering
\caption{\textbf{Evolution of RHEED patterns during MBE growth of AlN}. (\textbf{A} and \textbf{D}) Before growth, (\textbf{B} and \textbf{E}) after 5~nm growth and (\textbf{C} and \textbf{F}) by the end of growth of sample A (\textbf{A} to \textbf{C}) and sample B (\textbf{D} to \textbf{F}). All the RHEED patterns were taken along the AlN $<$11$\bar{2}$0$>$ azimuth.}
\end{figure}

\clearpage

\begin{figure}
\includegraphics[width=1 \textwidth]{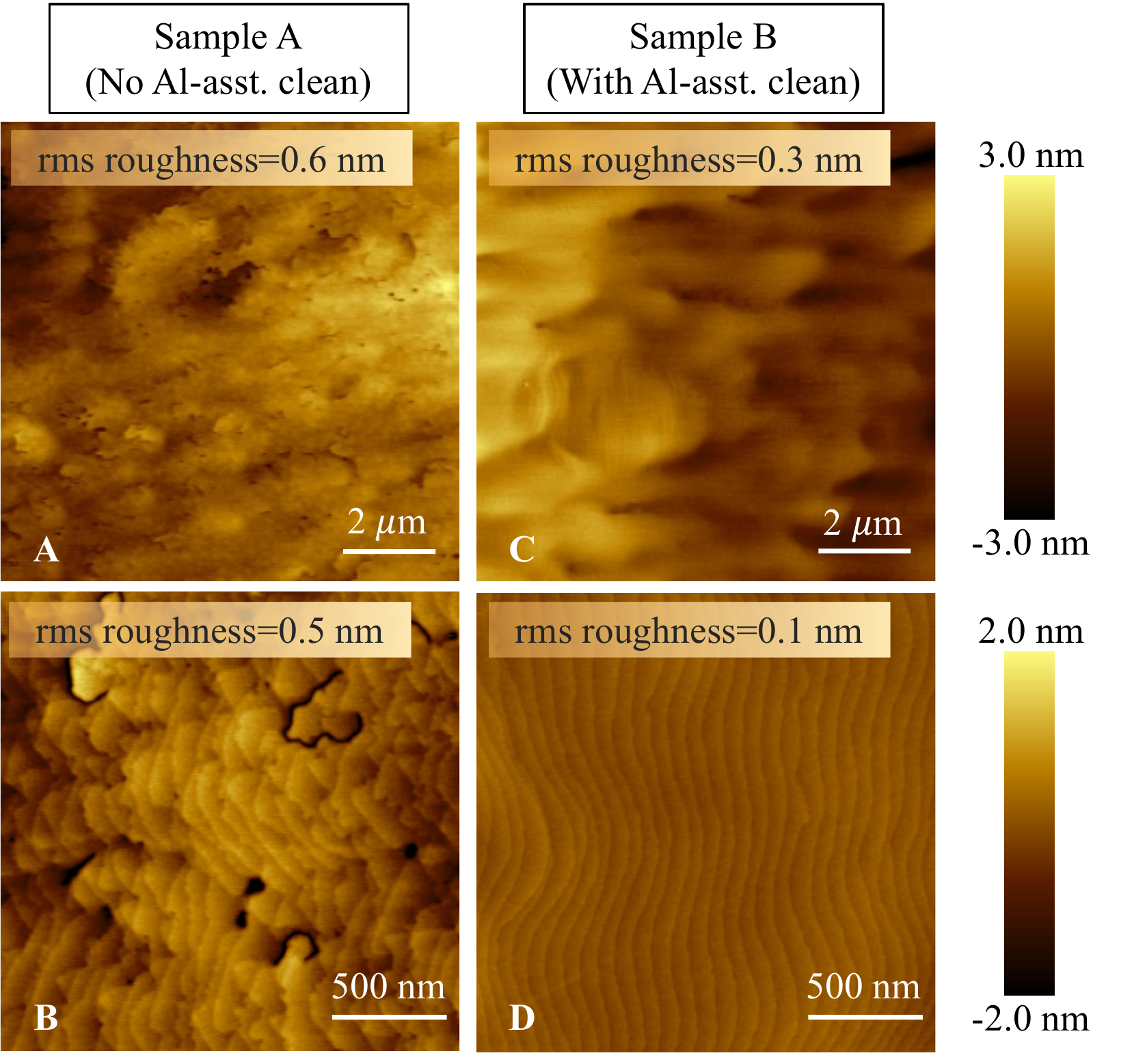}
\centering
\caption{\textbf{Morphology of as-grown surface of AlN}. (\textbf{A} and \textbf{C}) 10$\times$10~$\mu$m$^2$ and (\textbf{B} and \textbf{D}) 2$\times$2~$\mu$m$^2$ AFM micrographs of the surface of as-grown sample A (\textbf{A} and \textbf{B}) and sample B (\textbf{C} and \textbf{D}). Note that the pits and the trenches observed on sample A are absent on sample B.}
\end{figure}

\clearpage

\begin{figure}
\includegraphics[width=1 \textwidth]{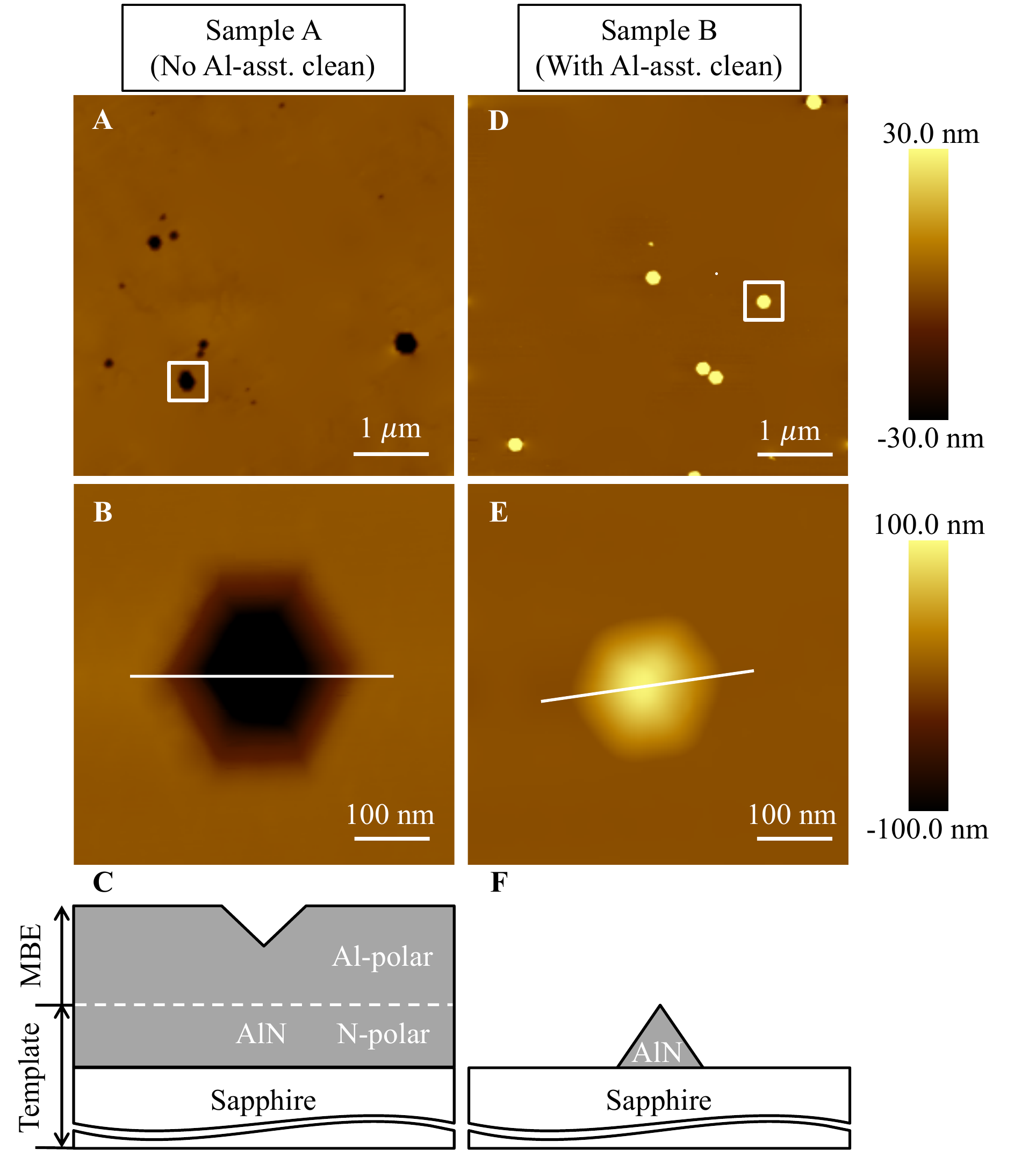}
\centering
\caption{\textbf{Surface morphology of AlN after KOH etch}. (\textbf{A} and \textbf{D}) 5$\times$5~$\mu$m$^2$ and (\textbf{B} and \textbf{E}) 0.5$\times$0.5~$\mu$m$^2$ AFM micrographs of sample A (\textbf{A} and \textbf{B}) and B (\textbf{D} and \textbf{E}) after KOH etch. Schematic of (\textbf{C}) sample A and (\textbf{F}) sample B after KOH etch. Note the hexagonal pits (\textbf{B}) on sample A and the pyramids (\textbf{E}) on sample B after KOH etch, which are signatures of Al-polar and N-polar AlN surfaces, respectively.}
\end{figure}

\clearpage

\begin{figure}
\includegraphics[width=0.8 \textwidth]{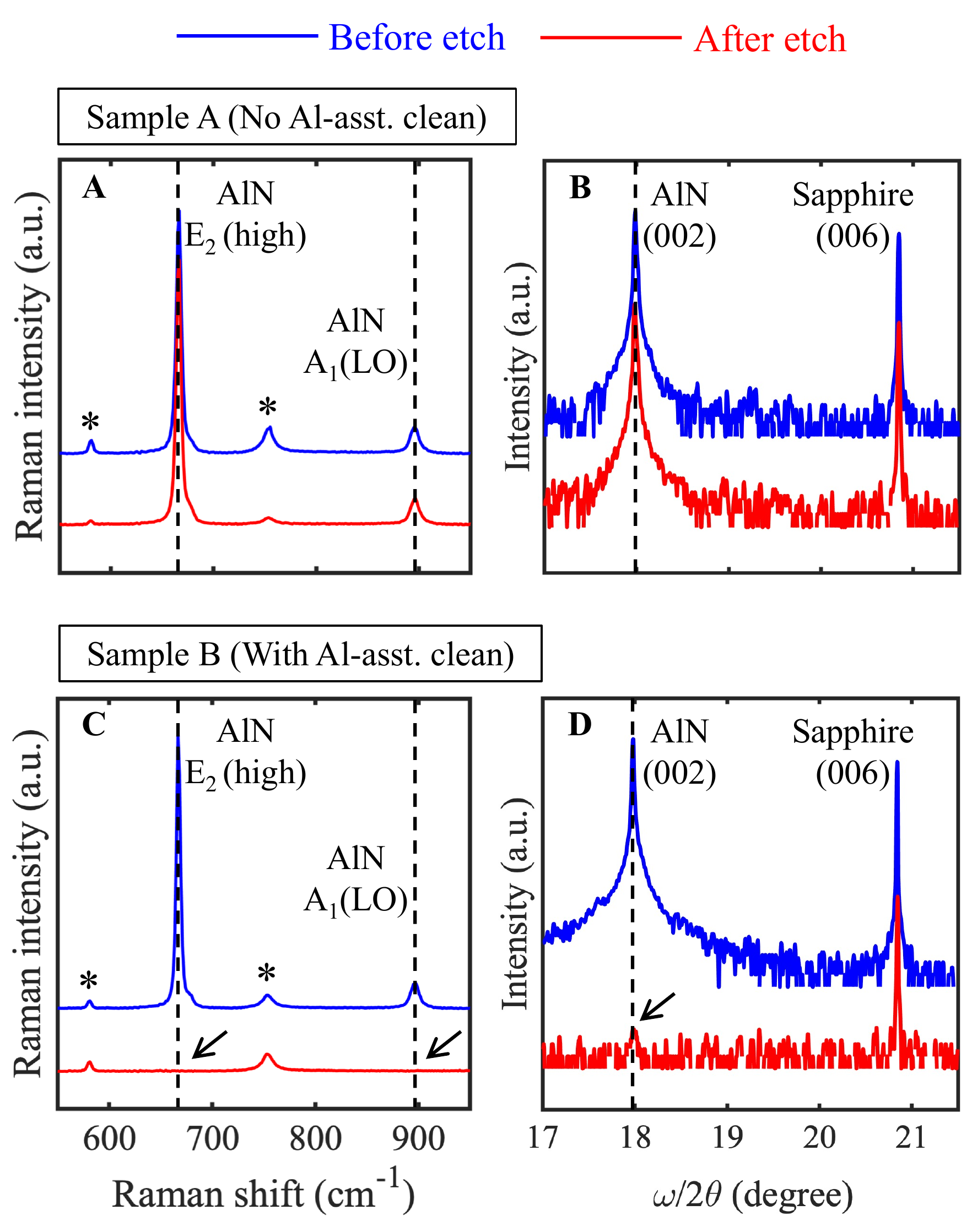}
\centering
\caption{\textbf{Comparison of XRD and Raman spectra before and after KOH etch}. (\textbf{A} and \textbf{C}) Raman spectra and (\textbf{B} and \textbf{D}) XRD $\omega/2\theta$ scans of sample A (\textbf{A} and \textbf{B}) and sample B (\textbf{C} and \textbf{D}) before (blue lines) and after (red lines) KOH etch. The black dashed lines in (\textbf{A}) and (\textbf{C}) indicate the AlN Raman modes. Raman signals from the sapphire substrate are marked by the asterisks. Note that the AlN signals in sample B are almost completely vanished after KOH etch, as indicated by the black arrows in (\textbf{C}) and (\textbf{D}).}
\end{figure}

\clearpage

\begin{figure}
\includegraphics[width= \textwidth]{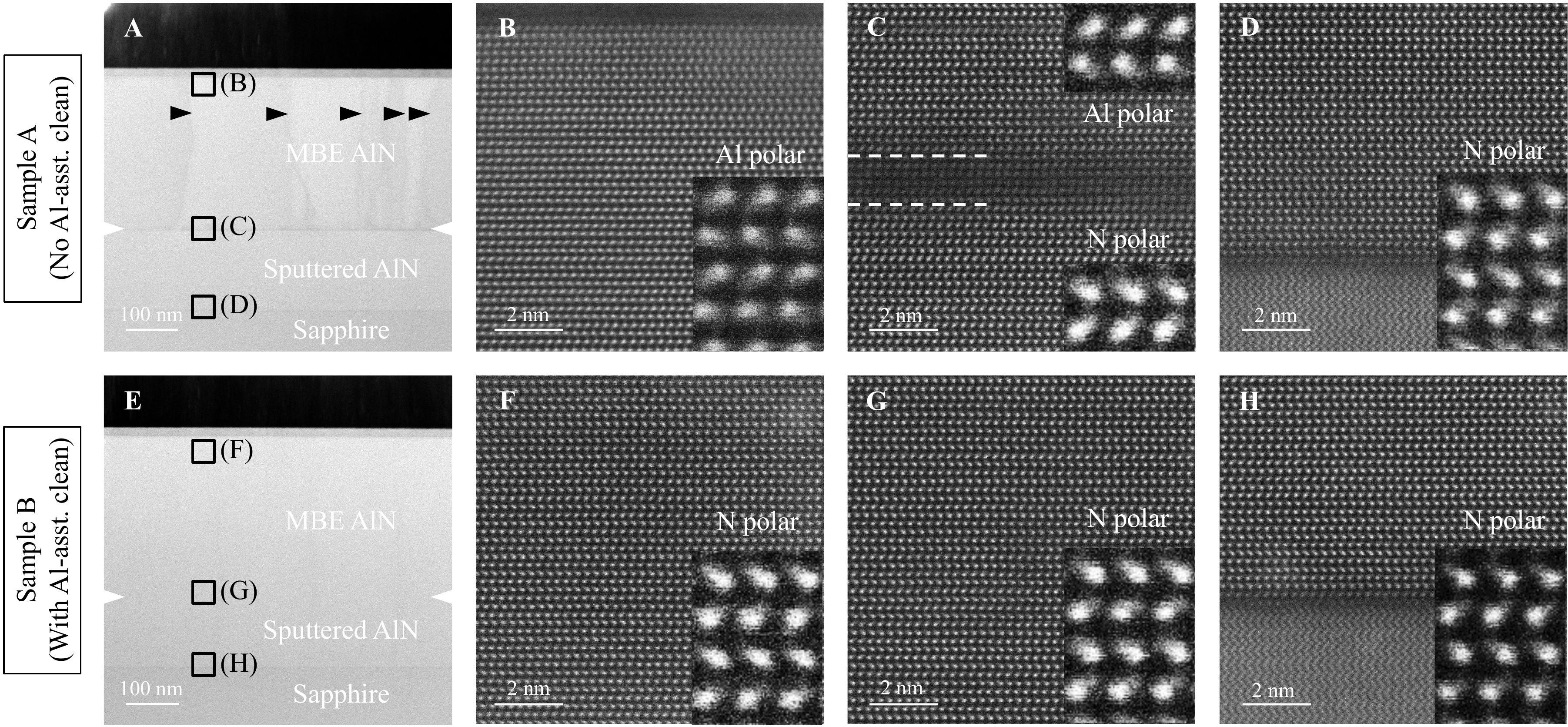}
\centering
\caption{\textbf{Cross-sectional STEM images of MBE-grown AlN on AlN templates}. (\textbf{A} and \textbf{E}) Overview images and (\textbf{B} to \textbf{D} and \textbf{F} to \textbf{H}) magnified HAADF-STEM images of sample A (\textbf{A} to \textbf{D}) and sample B (\textbf{E} to \textbf{H}). The black squares in (\textbf{A}) and (\textbf{E}) mark the regions where the corresponding magnified images (\textbf{B} to \textbf{D} and \textbf{F} to \textbf{H}) are taken. The white notches in (\textbf{A}) and (\textbf{E}) indicate the growth interfaces. Note the considerable image contrasts in the MBE layer and at the growth interface in sample A (\textbf{A}) are absent in sample B (\textbf{E}).}
\end{figure}

\clearpage

\begin{figure}
\includegraphics[width=1 \textwidth]{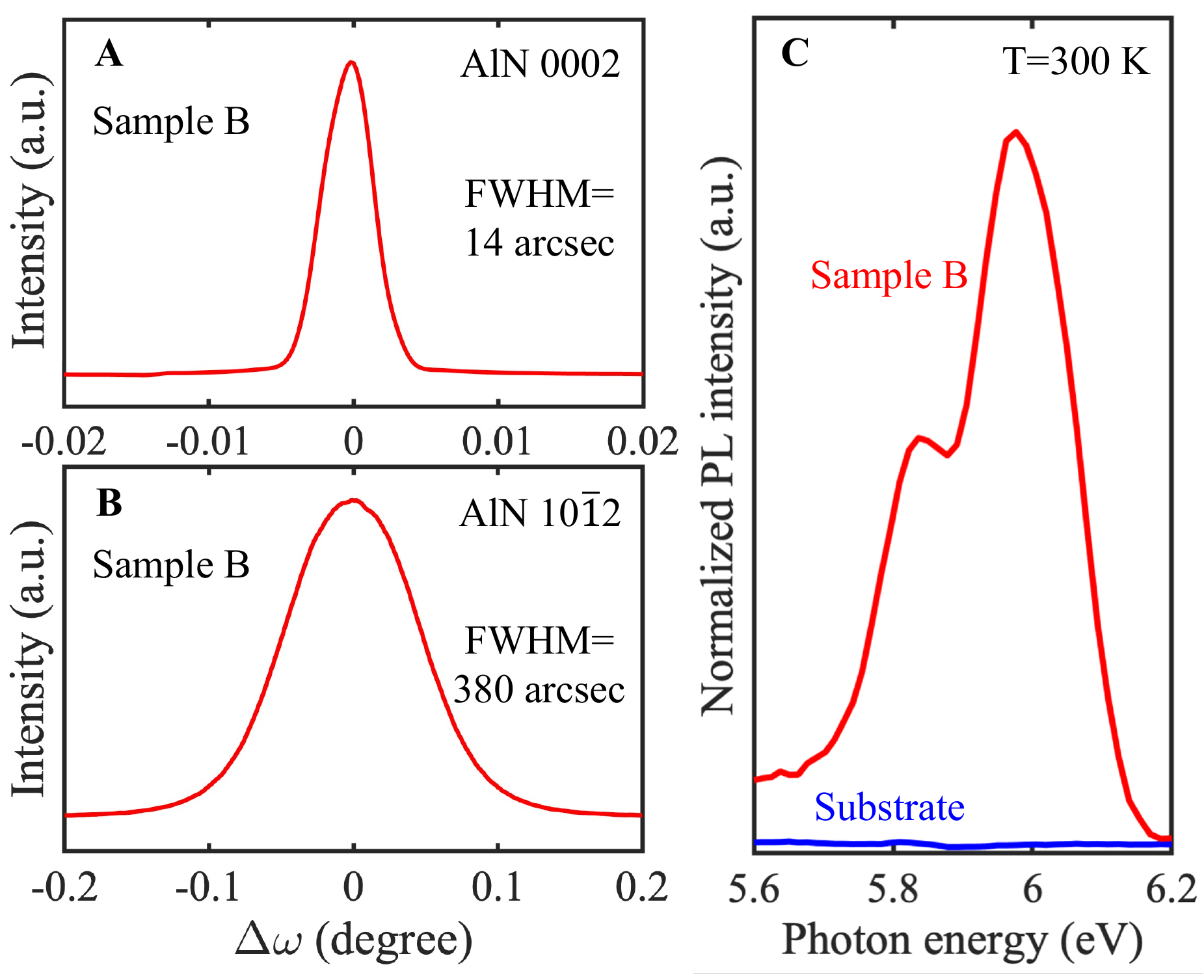}
\centering
\caption{\textbf{Structural and optical properties of MBE-grown N-polar AlN}. XRCs of sample B across AlN (\textbf{A}) (0002) and (\textbf{B}) (10$\bar{1}$2) reflections. (\textbf{C}) Room-temperature PL spectra around the band-edge of AlN of sample B (red line) and a bare AlN template (blue line). Note that near band-edge PL emission peaks are only observed on sample B, not on the bare AlN template.}
\end{figure}

\clearpage

\includepdf[pages=-]{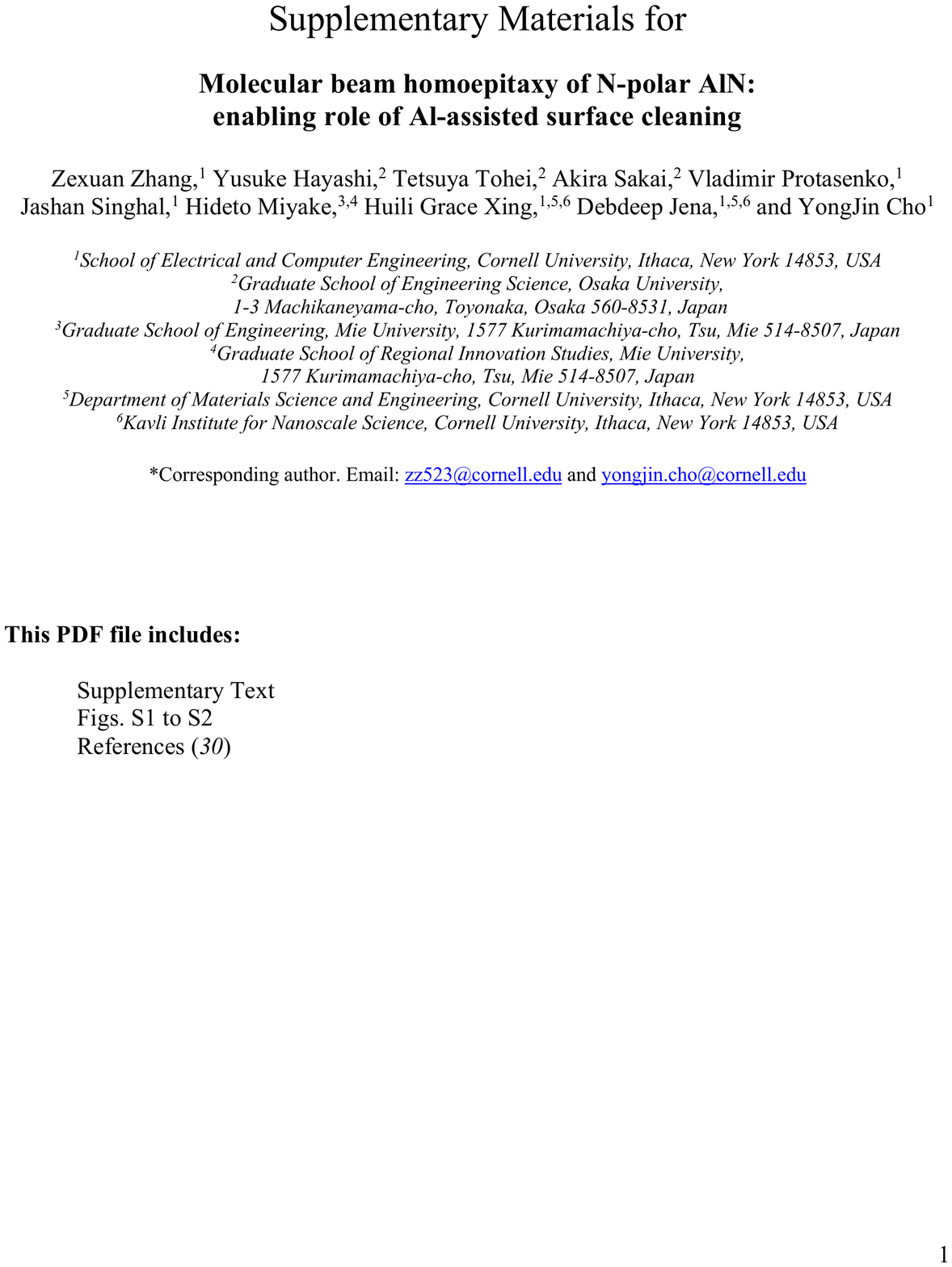}

\end{document}